\documentclass{aa}
\usepackage[varg]{txfonts}
\usepackage{amsmath}
\usepackage{amsfonts}
\usepackage{graphicx}
\usepackage{xcolor}
\usepackage[colorlinks=true,citecolor=blue,linkcolor=blue,urlcolor=blue,breaklinks=true]{hyperref} % to make refs linked to DOIs -- JvL   

\usepackage{aas_macros}
\usepackage{xspace}
\usepackage{natbib}%[square, comma, sort&compress]{natbib} % Use the natbib reference package - read up on this to edit the reference style; if you want text (e.g. Smith et al., 2012) for the in-text references (instead of numbers), remove 'numbers'
\bibpunct{(}{)}{;}{a}{}{,} % to follow the A&A style 

\def\refbf{}
\newcommand{\cand}{\mbox{J0337+62}\xspace}
\newcommand{\dm}{pc\,cm$^{-3}$\xspace}

\begin{document}

\title{A LOFAR search for steep-spectrum pulsars \\ in Supernova Remnants and Pulsar Wind Nebulae}
\author{S. M. Straal\inst{1,}\inst{2}\thanks{s.m.straal@uva.nl} \and J. van Leeuwen\inst{2,}\inst{1}}
\institute{Anton Pannekoek Institute for Astronomy, University of Amsterdam, Science Park 904, PO Box 94249, 1090 GE Amsterdam, The Netherlands
\and 
ASTRON, The Netherlands Institute for Radio Astronomy, PO Box 2, 7790 AA Dwingeloo, The Netherlands}
\titlerunning{Directed pulsar search to SNRs and PWNe with LOFAR}
\authorrunning{Straal \& van Leeuwen}
\date{Accepted 16 January 2019}

\abstract{Pinpointing a pulsar in its parent supernova remnant (SNR) or resulting pulsar wind nebula (PWN) is key for understanding its formation history, and the pulsar wind mechanism. Yet, only about half the SNRs and PWNe appear associated with a pulsar. We aim to find the pulsars in a sample of eight known and new SNRs and PWNe.
Using the LOFAR radio telescope at 150\,MHz, each source was observed for 3 hours.
We covered the entire remnants where needed, by employing many tied-array beams to tile out even the largest objects. For objects with a confirmed point source or PWN we constrained our search to those lines of sight.
We identify a promising radio pulsar candidate towards PWN G141.2+5.0. 
The candidate, PSR~J0337+61, has a period of 94\,ms and a DM of 226 \dm. 
We re-observed the source twice with increased sensitivities of 30\% and 50\% but did not re-detect it.
It thus remains unconfirmed. 
For our other sources we obtain very stringent upper limits of $0.8-3.1$\,mJy at 150\,MHz. 
Generally we can  rule out that the pulsars travelled out of the remnant. From these strict limits we conclude our non-detections towards point-sources and PWNe are the result of beaming and propagation effects. 
Some of the remaining SNRs \refbf{should} host a black hole rather than a neutron star. } %needs to be one paragraph

\maketitle

\section{Introduction}
\label{sec:introduction}
Finding young pulsars in their supernova remnant (SNR) is important for studying pulsar formation and the supernova explosion mechanism. The most energetic pulsars, ($\dot{\rm{E}} \gtrapprox 10^{36}$\,erg s$^{-1}$), produce a wind of highly relativistic particles which results in a pulsar wind nebula (PWN).
In PWNe the presence of a pulsar is beyond doubt. However, in only about \emph{half} of the known SNRs and PWNe are pulsars seen \citep{pwncatalog,green2014}.

\refbf{Neutron stars (NS) are formed} in core-collapse supernova remnants (cc-SNR).
For solar metallicity environments, 
the  fraction of such supernovae that produce a NS
lies between 75\% and 87\%
-- depending on the used lower mass limit for fall-back black-hole formation \citep{heger2003}.
\refbf{These young NSs come in different forms.
They can be observed in the high-energy band as central compact objects (CCOs) -- a class of weakly magnetised neutron stars unable to produce non-thermal emission or a wind nebula (see e.g. \citet{deluca2017} for a recent review).
Alternatively they can be highly magnetized and observable as magnetars.
But if the majority of young NSs emit in the radio as rotation-powered pulsars \citep{keane2008}}, 
the lack of detections of associated pulsars can be explained in two ways.
Firstly, missing these sources may follow from some  fundamental hurdles --  the beam of the pulsar might not be pointed towards earth, causing us to miss the pulse.
Secondly, instrumental limitations may prevent us from detecting these pulsars.  

Previous pulsar searches towards SNRs and PWNe have mostly been conducted with single-dish telescopes \refbf{using one or more beams}, at central frequencies around 600 and 1400 \mbox{MHz}, to counter the propagation effects of the Galactic plane. 
These \refbf{systems} have field-of-views (FoVs) smaller than many SNRs and PWNe. 
They are thus forced to observe in multiple epochs, of shorter integration times. 
This results in overall less sensitive observations. To reach the required sensitivity, observers are often forced to focus only on the central area of the SNR, causing them to miss any pulsars that might have traveled outside the FoV.
Now, the LOw Frequency ARray (LOFAR), has a large instantaneous FoV, allowing us to cover even the largest SNRs in a single observation \citep{2013A&A...556A...2V}. 
LOFAR is an interferometric radio array, 
spread out over Europe and operating between 10 and 250\,\mbox{MHz}. 
The Low Band Array (LBA) covers the frequencies from 10 to 90\,\mbox{MHz} while the High Band Array (HBA) can observe between 110 and 250\,\mbox{MHz}. 

Observing in this long-wavelength range can be a boon.
Pulsars are known to have steep flux-density spectra, 
making them  brighter at low frequencies.
When expressed as a power law
$\rm{S}_\nu \propto \nu^\alpha$, 
previous studies have obtained average spectral indices of $\alpha = -1.6\pm0.3$ \citep{lorimer1995}, $\alpha=-1.8\pm0.2$ \citep{maron2000}, and $\alpha=-1.4\pm1$ \citep{bates2013}. 
These are, however, obtained from observations well above 100\,MHz.
%which makes it tricky to use these indices to extrapolate pulsar flux densities to lower frequencies. 
From pulsar flux densities in the $102-400$\,MHz range, \cite{malofeev2000} obtain a mean spectral index of $\alpha=-1.47\pm0.76$ \refbf{and more recently, \citet{bilous2016} find a mean spectral index of $\bar{\alpha}=-1.4$ for non-recycled pulsars observed with LOFAR around 150\,MHz.}
In our study we would like to compare our sensitivity estimates to those of previous searches and we adopt $\alpha=-1.4$ which seems to correspond well to spectral slopes at higher and lower frequencies \citep{malofeev2000,bates2013,bilous2016}.

A further advantage is that normal pulsars have a spectral turnover around 130\,MHz and hence their flux density 'peaks' at these low frequencies \cite[cf.][]{izvekova1981,malofeev2000}.
This makes the 110$-$180\,MHz band well suited to search for pulsars.

Finally, at low frequencies more pulsars may be beamed towards us.
As lower frequencies are emitted higher up in the pulsar magnetosphere, 
pulsar beams are wider at lower frequencies, thus increasing the chance of a detection.  
Such widening of pulse profiles at low frequencies was confirmed for the LOFAR sample by \citet{pilia2016}.

Downsides to low-frequency observing are equally present, in the form of elevated sky background noise and increased pulse scattering. Yet, the benefits of the brighter emission at low radio frequencies, and the broader beams, may outweigh these, and increase the overall chance of detection.

We observed eight known and new SNRs and PWNe at low radio frequencies to search for associated pulsars. 
In Section \ref{sec:sources} we describe our source selection. Our observations and data reduction are described in Section \ref{sec:observations} and we present our results in Section \ref{sec:results}. 
A candidate detection prompted re-observations which we describe in Section \ref{sec:confirmation}. 
We discuss our results in Section \ref{sec:discussion} and conclude in Section \ref{sec:conclusions}.

\section{Source background and selection}
\label{sec:sources}
We selected the SNRs and PWNe that have no associated pulsar or black hole and are likely the result of a cc-SN. 
\refbf{We also include sources with a CCO as this class might be radio-faint rather than radio-quiet.}
We selected those sources that have ages $\lessapprox$ a few times $10^5$ years as \citet{gaensler1995_275}  find that for systems up to that age associations may still exist.
Only those targets were selected that are within the part of the sky where LOFAR is most sensitive (declination > 10\,deg.).
By next limiting our search to relatively nearby (d $\lessapprox$ 6\,kpc) sources,
the expected propagation effects on the pulsar signal are within the range of observing capabilities of LOFAR. 
\refbf{For that range, the dispersion can be accommodated by the channel width (cf.~\S\ref{sec:confirmation});
and as new pulsars are detected out to that distance in the LOFAR LOTAAS survey \citep{cooper2017},
the flux densities should be well within  range of the sensitivity attained in our 3$\times$ longer integrations (demonstrated in more detail in \S\ref{sec:lum}).
}

Based on these criteria, targets were selected from the online SNR catalog\footnote{online SNR catalog; \url{http://www.physics.umanitoba.ca/snr/SNRcat}}, \citep{ferrand2012} and Green's SNR catalog\footnote{Green's SNR catalog; \url{http://www.mrao.cam.ac.uk/surveys/snrs/}} \citep{green2014}.
Below we describe the selected sources, and relevant existing research, including previously conducted pulsar searches. 
\refbf{Parameters are quoted as given in the cited work.}

\begin{table*}
\begin{center}
\caption{Target parameters.}
\label{table:sources}
\begin{tabular}{l l c l l}
\hline\hline
Source & Other name & Size    & d     & Object\\
       &            & (arcmin)& (kpc) &       \\
\hline
G49.2$-$0.7 & W51C  & $\sim$30      & 5.4   & SNR (+ PWN) \\
G63.7+1.1   &       & 8       & $5.8 \pm 0.9$ & SNR + PWN        \\
G65.3+5.7   & G65.2+5.7 & 310 $\times$ 240 & 0.9 & SNR     \\
G74.9+1.2   & Cygnus Loop & 8 $\times$ 6 & $6.1\pm0.9$ & SNR + PWN\\
G93.3+6.9   & DA530 & 27 $\times$ 20 & 3.5 & SNR (+PWN)          \\
G141.2+5.0  &       & 3.5     & $4.0 \pm 0.5$ & PWN        \\
G150.3+4.5  &       & $\sim$150 &    & SNR          \\
G189.22+2.9 & IC443 & 1\farcs4 & 1.5 & SNR + PWN + NS    \\
\hline
\end{tabular}
\end{center}
%\tablefoot{*The distance is given in arcseconds rather than arcminutes}
\end{table*}

\subsection{G49.2--0.7}
\label{sec:G49}
Also known as W51C, G49.2$-$0.7 is a middle-aged SNR \citep[$\sim 30$\,kyr,][]{koo1995}. 
It lies in the  W51 complex that also consists of \ion{H}{ii} region W51A and the massive-star forming region W51B \citep{tian2013}.

Previously, \citet{tian2013} and \citet{koo1995} had placed the SNR at a distance of respectively 4.3\,kpc and $\sim$6\,kpc. 
This has recently been revised to 5.4\,kpc \citep{ranasinghe2018}, which we adopt in this work.
At this distance, the angular diameter of the source \citep[$\sim$30\arcmin;\,][]{copetti1991} translates to a physical size of 47\,pc.

\cite{koo2002} observed the complex with \textit{ASCA} and identified a hard X-ray source, CXO J192318.5+1403035, as a candidate PWN. 
Further \textit{Chandra} observations by \cite{koo2005} show that the structure of the hard X-ray source consists of a diffuse envelope and a core containing a compact source, presumably the powering pulsar.

G49.2$-$0.7 has previously been searched by \citet{gorham1996} with Arecibo at 430\,MHz and 1420\,MHz. Yet only the central 15$\%$ and 2$\%$ of the remnant were respectively covered and searched down to minimum sensitivities ($S_{\rm{min}}$) of 0.6\,mJy and 0.5\,mJy.
This search did not cover the hard X-ray source.

The location of G49.2$-$0.7 in a star-formation region makes it a likely cc-SNR and  thus a good target to search for a possible pulsar association.

\subsection{G63.7+1.1}
\label{sec:G63}
G63.7+1.1 was discovered by \cite{taylor1992} and confirmed as a plerionic (filled-centre) SNR by \cite{wallace1997}. The latter also determined the kinematic distance to the source to be $3.8\pm1.5$\,kpc, based on associated \ion{H}{i} and CO features.

Recently, G63.7+1.1 was observed in X-rays for the first time by \cite{matheson2016} \refbf{with \textit{XMM Newton} and \textit{Chandra}}. \refbf{They aimed to detect the PWN and searched for the associated neutron star.} 
\refbf{\citet{matheson2016} find clear evidence for an X-ray nebula and suggest that the CXO J194753.3+274351, located near the peak of the X-ray emission of the remnant, is the powering neutron star.}
\refbf{Additionally, }\cite{matheson2016} provide an updated distance to the source of $5.8\pm0.9$\,kpc. The diameter of the radio nebula of 8\arcmin~then translates to a physical size of 14\,pc. Assuming the pulsar was born at the peak of the radio luminosity and has traveled to the location of the X-ray peak, applying the average observed 2D pulsar speed of 246 $\pm$ 22\,km s$^{-1}$ \citep{hobbs2005} leads to an age estimate of roughly $\sim$8\,kyr.

Even though \cite{matheson2016} provide estimates for the putative pulsar energetics, no search for an associated pulsar has been reported.

\subsection{G65.3+5.7} 
\label{sec:G65}
G65.3+5.7 (also known as G65.2+5.7) is a large SNR, with major axes of $310\arcmin \times 240\arcmin$,  discovered by \cite{gull1977}.
The distance to the remnant, derived from the velocity of optical line emission, is $\sim 0.8 - 1.0$\,kpc \citep{lozinskaya1981}. That is consistent with a previous estimate of $0.9\substack{+0.6 \\-0.3}$\,kpc based on the diameter and surface brightness ($\Sigma-D$) relation by \cite{reich1979}. 
Both distances agree well with that expected from its X-ray luminosity \citep{schaudel2002}. 
At this distance the physical size is estimated to be $75\substack{+50 \\-25}$\,pc, at an age of less than $\sim2.4\times10^5$ yr.

A radio pulsar search at 1420\,MHz by \cite{gorham1996} covered the inner 3\arcmin of the remnant, and did not detect any pulsar down to a sensitivity of 0.1\,mJy. 
However, this non-detection is not very constraining as a transverse velocity of $v_{\rm{trans}} = 20$\,km/s would be sufficient to move the neutron star out of the searched area.
Outside the searched area by \cite{gorham1996}, \cite{camilo1996} found a 0.58\,s pulsar at 45' from the center of the remnant with a flux density of 1.9\,mJy at 606\,MHz. 
This pulsar (J1931+30) seems too slow-spinning to be associated with the remnant. \refbf{It is possible that a slow-spinning pulsar is formed, however} its dispersion measure ($56\pm5$ \dm) suggests a distance of 3.4\,kpc (NE2001; \citealt{cordes2002}), three times further than the  remnant. As the remnant is located in a low ambient matter density environment \citep{schaudel2002}, such a DM discrepancy suggests the pulsar is  not  associated with the SNR.
 G65.3+5.7 was part of the sample of six large-diameter SNRs in which \cite{kaplan2006} aimed to identify X-ray point sources from the \textit{ROSAT} Bright Source Catalog in and near the SNRs to find the associated neutron star. Down to a luminosity of $\sim 10^{32}\,\rm{ergs\,s^{-1}}$ no associated neutron star was found.

\subsection{G74.9+1.2}
\label{sec:G74}
G74.9+1.2 is a plerionic SNR, also known as CTB\,87. 
\citet{matheson2013} reveal a compact X-ray nebula, 100\arcsec~offset from the radio peak. The nebula shows a torus/jet-like structure, which indicates its origin as a pulsar wind nebula. The putative pulsar would then be located at the detected point source CXOU J201609.2+371110. The observed properties in \cite{matheson2013} suggest that the source has an age of $\sim5-28$\,kyr.

\cite{kothes2003} show that a previous distance determination of 12\,kpc \citep{green1989,wallace1997} is erroneous due to the assumption of a flat Galactic rotation curve in the direction of G74.9+1.2.
By applying the extinction-distance relation introduced by \cite{foster2003} the distance is adjusted to 6.1 $\pm$ 0.9\,kpc. That is in excellent agreement with distance determinations using the thermal electron model of \cite{taylor1993}.
The remnant size of 8\arcmin $\times$ 6 \arcmin \citep{green2009} implies a physical size of $\sim$14$\times$11\,pc at a distance of 6.1\,kpc.

Although \cite{matheson2013} detect an X-ray point source, previous searches have not found a radio pulsar. \cite{gorham1996} fully searched G74.9+1.2 using the Arecibo radio telescope at 430\,MHz and reached a sensitivity of 0.4\,mJy. 
\cite{biggs1996} searched the source at 606\,MHz using the 76-m Lovell radio telescope. They did not find a pulsar down to a flux density of 10\,mJy.
The source was again searched, but deeper, by \cite{lorimer1998} with the 76-m Lovell radio telescope at 606\,MHz with an 8\,MHz bandwidth. They obtain an upper limit of $S_{\rm{min}}$ = 1.1\,mJy.

\subsection{G93.3+6.9}
\label{sec:G93}
G93.3+6.9, also known as DA530, is one of the rare SNRs observed at high Galactic latitudes. In Green's SNR catalog only seven out of 256 Galactic SNRs have $|$b$|$ $>$ 6$^\circ$. The other six SNRs are large-diameter nearby SNRs. G93.3+6.9 on the other hand, is a well-known example of a sub-energetic SNR. The SNR has a well-defined shell-like radio morphology and is extremely faint in X-rays \citep{landecker1999}. X-ray spectral models indicate that the SN occurred in a low-density medium, consistent with a wind-blown bubble \citep{kaplan2004} typical for massive stars, indicating a cc-SN. 
From the further lack of enhanced iron lines, \citet{jiang2007} argue that the SNR is likely the product of a core-collapse SN of an 8-12 M$_{\sun}$ progenitor. A small-scale hard X-ray feature is detected near the center of the remnant which \citeauthor{jiang2007} hypothesize could be a PWN. 
\refbf{\citet{jiang2007} conclude that the stellar remnant must be a neutron star based on the subenergetic SN and the presence of a PWN.}
The Sedov blast model suggests a dynamical age for G93.3+6.9 of $\sim$ 5 $-$ 7\,kyr.

The angular diameter of 27\arcmin~$\times$ 20\arcmin~\citep{roger1976} translates to a physical size of $17 \times 13$\,pc at a distance of $2.2\pm0.5$\,kpc \citep{foster2003}.  

\cite{lorimer1998} carried out a radio-pulsar search using the 76-m Lovell radio-telescope at 606 MHz with an 8\,MHz bandwidth. 
The remnant is larger than the telescope FoV, forcing observations to be split in  multiple 34-minute pointings.
This setup reached an estimated minimum flux density of $S_{\rm{min}}$= 0.8\,mJy. No pulsations were found. An X-ray study of the source by \citet{kaplan2004} shows no signs for an associated compact X-ray source. Little is yet known about sub-energetic SNRs and the compact objects they produce.

\subsection{G141.2+5.0}
\label{sec:G141}
G141.2+5.0 is a newly found PWN \citep{kothes2014}. 
It has a diameter of $\sim$3\farcm5, which  at a distance of 4.0\,kpc translates to a physical size of $\sim$4\,pc. 
\citet{kothes2014} also reveal a small \ion{H}{i} bubble enclosing the PWN, which could be either the parent SNR or a progenitor-blown wind-bubble. 
G141.2+5.0 is the first SNR/PWN found in the Cygnus spiral arm and has a steep spectrum, which makes it part of a rare class of steep spectrum PWNe. 
In one of the other sources in this class, G76.9+1.1, the steep spectrum was not caused by the usual aging, but a highly energetic, young pulsar was found to power it.
This suggests a young nature also for our target.
\refbf{No age estimate is provided.}

Recently, a compact X-ray source has been found at the center of G141.2+5.0, confirming its PWN nature \citep{reynolds2016}. 
A search for X-ray pulsations was limited to  a time resolution of 3.141\,s. That only allows for detection of  periods beyond 12.5\,s, which is slower than the rotation period for all but one known isolated neutron stars.
A search for the putative pulsar with the Green Bank Telescope (GBT)\refbf{, encompassing the entire PWN,} at 820\,MHz, with a 200\,MHz bandwidth for 6,000\,s, did not result in any detections (D.~Lorimer, \emph{priv.~comm.}).

\subsection{G150.3+4.5}
\label{sec:G150}
This SNR is recently discovered by \cite{gao2014} in the Urumqi $\lambda$6\,cm Galactic plane survey data.
G150.3+4.5 is an elliptical (2$\fdg$5 $\times$ 3$\degr$) shell-type SNR with a typical steep radio synchrotron spectrum of $\alpha= -0.6$.
Part of the shell overlaps with SNR G149.5+3.2. 

The 2FHL \textit{Fermi} catalog \citep{fermi2016} reports on a high-energy counterpart to G150.3+4.5. 2FHL\,J04031.2+5553e is a hard-spectrum, extended source and is spatially coincident with G150.3+4.5. \citet{ackermann2017} search for extended sources within 7$\degr$ of the Galactic plane using 6 yrs of \textit{Fermi-}LAT data. 
The hard spectrum of G150.3+4.5 in the 10\,GeV -- 2\,TeV band is similar to that of young shell-type supernova remnants, but its large size and faintness suggest it is old.
\refbf{Finding a young pulsar would hence be interesting}.
The age of the remnant remains inconclusive.  

Since this source is only recently identified as a SNR, \refbf{no pulsar searches have been reported. The unknown distance cannot act as a criterion to exclude this source. As its location off the Galactic plane forecasts relatively low dispersion, we include it in our sample. For completeness we search out 
to a DM of 400, above which we expect difficulties in finding a signal at our low observing frequency. As this is twice the maximum DM expected for this entire line of sight \citep{cordes2002,ymw16}, the source dispersion measure falls within our search capabilities. }

\begin{table*}[h]
\caption{
Observational set-up for each of our targets. 
The given coordinates show the center of each pointing. 
The overlap indicates the FWHM fraction overlap of the tied-array beams (TABs). 
See Fig.~\ref{fig:beams} for a representation of the observing set-up.}
\label{table:observations}
\centering
\begin{tabular}{l l l c c c c c c}
\hline\hline
Source & RA & Dec & Obs. date  & \# TABs & Beamsize & Overlap & BW & $\nu_\mathrm{c}$ \\
       & (J2000) & (J2000) & yyyy--mm--dd & & (arcmin) & (FWHM) & (\mbox{MHz}) & (\mbox{MHz}) \\
\hline
G49.2$-$0.7 &  19h23m19\fs 2 & $+14\degr 09\arcmin 31\arcsec$ & 2015--04--02 & 89* & $4\farcm$5 & 1 & $68.55$ & 146.00 \\
G63.7+1.1 & 19h47m56s & $+27\degr 44\arcmin 48\arcsec$ & 2015--04--18 & 7 & $4\farcm$5 & 0.5 & 78.125 & 149.61 \\
G65.3+5.7 & 19h34m00s & $+31\degr 12\arcmin 00\arcsec $ & 2015--04--16 & 91 & 30\arcmin & 0.9 & 78.125 & 149.61  \\
G74.9+1.2 & 20h16m09\fs 2 & $+37\degr 11\arcmin 10\arcsec$ & 2015--03--25 & 1 & 4\farcm 5 & - & 78.125 & 149.61 \\
G93.3+6.9 & 20h52m19s & $+55\degr 20\arcmin 00\arcsec$ & 2015--03--25 & 88* & 4\farcm 5 & 1 & 68.55 & 146.00 \\
G141.2+5.0 & 03h37m12s & $+61\degr 53\arcmin 05\arcsec$ & 2015--04--08 & 7 & 4\farcm 5 & 0.5 & 78.125 & 149.61 \\
G150.3+4.5 & 04h27m09s & $+55\degr 27\arcmin 30\arcsec$ & 2015--04--11 & 91 & 30\arcmin & 0.5 & 78.125 & 149.61 \\
G189.1+3 & 06h17m05\fs 18 & $+22\degr 21\arcmin 27\farcs 6$ & 2015--08--29 & 1 & 4\farcm 5 & - & 78.125 & 149.61 \\
\hline
\end{tabular}
\tablefoot{*For two beams toward G49.2$-$0.7 and three toward G93.3+6.9 recording failed.  }
\end{table*}

\subsection{G189.1+3}
\label{sec:189}
G189.1+3 is better known as SNR IC\,443, and it hosts PWN G189.22+2.90, discovered by \cite{olbert2001} using \textit{Chandra} and VLA observations. 
We will refer to the entire object as G189.1+3.
The nebula shows a cometary morphology and hosts an unresolved point source (CXOU\,J061705.3+222127) near its apex. 
This soft X-ray point source is probably the young pulsar powering the nebula and is well-observed.
\cite{gaensler2006} provide support for a physical association of the PWN and the SNR. Amongst others, they show that the thermal emission from CXOU\,J016705.3+222127 is consistent with standard cooling at the estimated age of 30\,kyr \citep{chevalier1999} for IC\,443. 
This is further supported by 
the  high spatial X-ray \textit{Chandra} spectra of G189.22+2.90 
obtained by \cite{swartz2015}, 
which are fit with  a hydrogen atmosphere model. 
From the derived effective temperature and bolometric luminosity, similar temperature and luminosities were obtained in some neutron star cooling models for $\tau \sim$ 30\,kyr.
This is in agreement with previous research on the remnants age. 
A search for X-ray pulsations, restricted to periods longer than 6.5\,s yielded no detection.

Early observations of the parent SNR, IC\,443, suggest an age of $\sim$3\,kyr \citep{petre1988}. However, \cite{chevalier1999} obtain an age of 30\,kyr for a physical size of 7.4\,pc. 

\citet{ambrocio2017} find a kinematic distance to the source of 1.9\,kpc. This corresponds to a radius of the SNR of 10.5\,pc. There the velocity of the filaments is explained by an age of the system of, again, 30\,kyr. Earlier work however assumes a distance of 1.5\,kpc as provided by \citet{welsh2003}. In this work we adopt a distance of 1.5\,kpc for consistency. 

The X-ray structure of the PWN is discussed in great detail by \cite{gaensler2006}. Its diameter of 1\farcs4 translates to a distance from the point source to the forward termination shock of 0.6\,pc. The surface temperature measured by \cite{gaensler2006} for the putative pulsar J0617+2221 is consistent with the aforementioned  age of 30\,kyr and given the consensus we adopt this age for the source.

The high-energy spectrum of the source and its flux are consistent with that of a rotation-powered pulsar \citep{swartz2015}. 
But previous searches could not turn up a pulsar. \cite{manchester1985} searched the remnant, \refbf{including the later discovered point source,} at 1.4\,GHz down to a sensitivity of 1\,mJy and \cite{kaspi1996} provide upper limits of 1.5\,mJy at 436\,MHz and 0.4\,mJy at 1520\,MHz\refbf{, where at least one of these observations covered the entire remnant and the other focused at its centre}.

\begin{figure*}
\centering
\includegraphics[width=\textwidth]{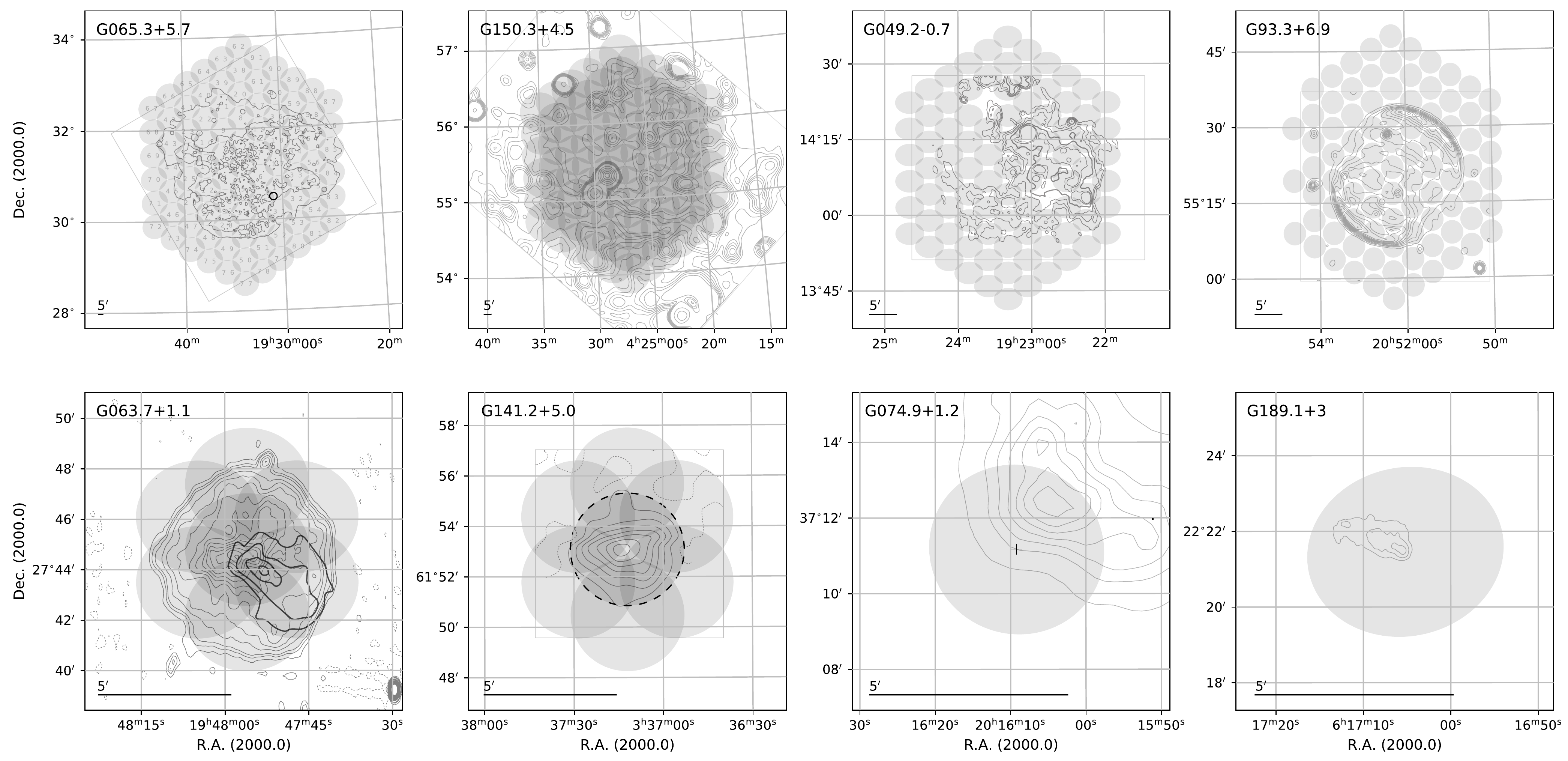}
\caption{
	The beam pattern for the 8 observations, ordered from largest (top left) to smallest diameter (bottom right). Observations in the top row all employ 91 tied array beams. In the bottom row, patterns contain either 7 (left) or 1 (right) beams. The general beam numbering pattern is visible in the top-left subpanel.
    For the respective sources, the background contours (gray) and observation bounding boxes (light gray) are derived from:
    G65.3+5.7:  ROSAT 0.44$-$1.21\,keV \citep{2004ApJ...615..275S}. Pulsar J1931+30 position marked by black  circle ---
    G150.3+4.5: Effelsberg 1.4\,GHz    \citep{gao2014}  ---
	G49.2-0.7:  VLA 1.4\,GHz           \citep{koo2005}  ---
	G93.3+6.9:	DRAO Synthesis Telescope 1.4\,GHz  \citep{landecker1999} ---
	G63.7+1.1:  Westerbork Synthesis Radio Telescope 
                (WSRT) 1.4\,GHz \citep{wallace1997} in gray; \refbf{X-ray nebula  \citep{matheson2016} in black}  ---
	G141.2+5.0:	WSRT 327\,MHz          \citep{kothes2014}. Here our candidate-confirmation pointing is marked with a dashed ellipse. ---
    G74.9+1.2:  VLA 1.4\,GHz   \citep{2018arXiv180409978L}. The X-ray nebula identified by \cite{matheson2013} is marked with a cross. ---
    G189.1+3.0: VLA 8.5\,GHz    \citep{gaensler2006}.
\label{fig:beams}
}
\end{figure*}

\section{Observations and data reduction}
\label{sec:observations}
\subsection{Observations}
We targeted our observations to cover the above mentioned sources in their entirety, unless a PWN or point source had been confirmed. In those cases we only observed the PWN or point source.
The observations were obtained using the High-Band Antennas (HBA) operating in the 110$-$188\,MHz range (project code LC3$\_$024). 
\refbf{We used the  Core stations, which make up the inner 2-km radius of LOFAR. Up to 24 of these are available. As they operate on the same clock signal, their data can be immediately tied-array beam formed \citep[][]{stappers2011}.}
For each observation we tiled out the area source using tied-array beams (TABs). 
As each source differs in size (see Table \ref{table:sources}), the observational set up per source is customized. We specified each observation such that the entire area of the source would be covered while ensuring a maximum possible sensitivity. This entails maximizing the allowable data-rate and for the biggest sources only observe with the Superterp (inner 6) stations which produces a larger beam size.
The resulting beam patterns, including an occasional missing beam for G49.2$-$0.7 and G93.3+6.9, 
are detailed in Table \ref{table:observations} and
 shown in Fig.\ref{fig:beams}. 

For each source we recorded 3 hours of stokes I data with 32 channels per subband at 8 bits per sample and a time resolution of 1.3\,ms. This time-resolution is sufficient to detect young pulsars which are expected to have periods in the 10$-$100\,ms range. 
Additionally, 
the resulting channel widths of 6.1035\,kHz allow for little loss in sensitivity when reducing the data at higher DMs (see Sec. \ref{sec:datareduction}). 
To verify the system we recorded nearby pulsars that \cite{pilia2016} had shown to be well visible, before each observation, using the exact same search set-up.
To maximize sensitivity allowed by the fixed LOFAR HBA tiles, all observations were scheduled around transit.

\subsection{Data Reduction}
\label{sec:datareduction}
After each observation the data were pre-processed by the LOFAR pulsar pipeline \citep{2010ASPC..434..193A, stappers2011} and saved in {\tt PSRFITS} format \citep{hotan2004} on the LOFAR Long-Term Archive\footnote{LTA; \url{http://lofar.target.rug.nl/}}. 
These 30\,TB of data were transferred using GridFTP to the Dutch national supercomputer Cartesius\footnote{\url{https://userinfo.surfsara.nl/systems/cartesius}}.

We further reduced the data using PRESTO\footnote{\url{http://www.cv.nrao.edu/~sransom/presto/}} \citep{ransom2001}. The data were cleaned from RFI and searched for periodic and single pulses. 
To reduce the computational cost, we limited the dispersion search range of each source, \refbf{based on the DM predicted by the \citet{cordes2002} and \citet{ymw16} electron density models (see Table~\ref{table:data_reduction}).}  
The data were de-dispersed in steps of 0.01 \dm up to a DM of 300 \dm and in steps of 0.03 \dm above. 

We inspected all candidates down to a PRESTO-reported $\sigma=4$, equal to a chi-squared of 1.8. This resulted in $\sim$450 candidates per beam, totalling a staggering 150,000 candidates. For each source we were able to blindly find the observed test-pulsar and/or a side-lobe pulsar which verified our observational set-up.

For the single pulse search we used the sifting algorithm developed by \citet{Michilli-2018b}. All candidates with $\sigma > 8 $ were inspected. 

\begin{figure}
\centering
\includegraphics[width=\columnwidth]{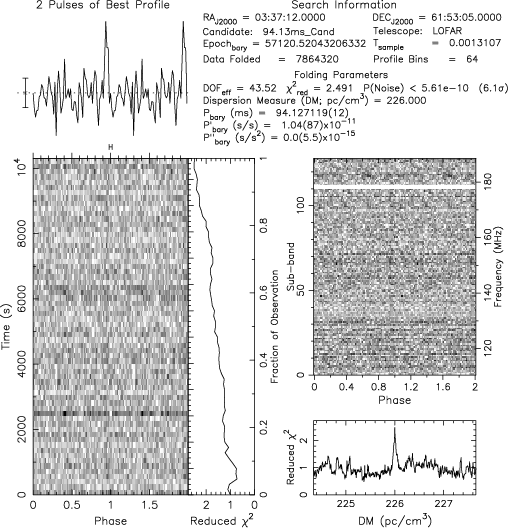}
\caption{
Candidate detection toward G141.2+5.0.
The signal is broadband and present throughout the
entire 3 hour observation. The DM is 226 \dm, consistent with the expected distance
and contribution of the nebula itself. Its period of
94.13\,ms is compatible with a young pulsar.
\label{fig:cand}
}
\end{figure}

\section{Results}
\label{sec:results}
\refbf{\subsection{Pulsar candidate towards G141.2+5.0}}
We detected a very interesting candidate in the central G141.2+5.0 pointing, shown in Fig.~\ref{fig:cand}. We hereafter refer to it as \cand.
Its position in the central beam 
is consistent with the position of the central X-ray point source \citep{reynolds2016}.
Its 94-ms period is as expected for a young pulsar.
The PRESTO pipeline ranked it as $\sigma=6$, and reports that the pulse profile deviates from a straight line at the $\chi^2=2.5$ level. That is a significance that inspires confidence: new LOFAR pulsars have been discovered with lower (less significant) values for  $\chi^2$. The second-found LOFAR pulsar, J0613+3731 \citep{coenen2014} was discovered at $\chi^2=2.4$ (Fig. 4.8 in \citealt{coen13}). Of the 83 pulsars discovered with LOFAR since\footnote{\label{fn:lotaas}\url{www.astron.nl/lotaas/}}, 11 have $\chi^2 \leq 2.4$; with detections as dim as illustrated by PSR~J1342+65, at $\chi^2=1.7$ and $\sigma=4.2$.

The DM of \cand,  226\,\dm, is
consistent with the expected DM plus a contribution from the nebula itself. At the proposed distance of $4.0\pm0.5$\,kpc on this line of sight, the dispersion measure from the ISM predicted by the NE2001
 model would amount to $123.21\substack{+14 \\-17}$ \dm  (cf.~Table~\ref{table:data_reduction}).
 \refbf{This implies that $\sim 40\%$ of the DM should be accounted to the direct environment.
 Although high, this is not unusual. 
 PSR J0908$-$4913 resides in a high density PWN \citep[> 2 \dm, see ][]{gaensler1998} and is observed through the Gum nebula, which together could contribute 100 \dm to the total DM \citep{cordes2016}.
}

If confirmed, pulsar J0337+62 would have the highest DM discovered by or otherwise seen with LOFAR \citep{bilous2016}.

Even after dedispersion the observed pulse should still contain the  imprint of multi-path scatter broadening, if it is astrophysical.
The observed pulse peak  indicates a maximum scattering time ($\tau_s$) of about 6\,ms (Fig.~\ref{fig:cand}).
The scattering time derived by \citet{kuzmin2007} for regular pulsars at 100\,MHz 
%these low frequencies 
roughly scales with DM as 
% \begin{equation}
$\tau_{s,\rm{100\,MHz}} = 60\times\left(\frac{\rm{DM}}{100}\right)^{2.2}\space\rm{ms}$.
% \end{equation}
However, as shown by \cite{cordes2016}, pulsars in PWNe can be severely under-scattered, or over-dispersed. 
\refbf{For a PWN pulsar such as J0908$-$4913 the dispersion measure}
needs to be reduced by $\sim130$\,\dm (which is by almost 75\%) to follow the DM -- scattering relation from \cite{cordes2016}.
In systems like J0908$-$4913, and probably in a number of the targets from the current work,
the higher gas density intrinsic to the PWN increases the final DM, but not the scattering.
Such local DM increases are also seen in the ensemble studies in \citealp{Straal-2018}.
Likewise we can here assume that the scatter-inducing, interstellar DM to the source follows  
from NE2001 (Table \ref{table:data_reduction}, DM$_{\rm{ISM}}=$\refbf{$123.21\substack{+14 \\-17}$}), and the remaining DM is local to the source, where it has little effect on the scatter broadening. 
%If for the scattering relation we start from the lower bound of 106\,\dm,
\refbf{If we determine the expected scattering following \citet{kuzmin2007}, using the lower bound ISM contribution to the DM of 106 \dm}
and we scale $\tau_{s}$ to 150\,MHz as $\nu^{-4.4}$,
the scatter broadening should be about 11\,ms. 
That is slightly higher than the observed width pulse, but within the order-of-magnitude scatter around the relation  visible in Fig.~3 of \citet{kuzmin2007}.

Even for this high DM, low scatter broadening, and modest signal-to-noise ratio (S/N), the broadband nature of the signal, visible in Fig.~\ref{fig:cand}, plus
the fact that the period, period derivative and dispersion measure are within the expected range for a young pulsar, drove us to classify this candidate as interesting for follow-up, as described in Section \ref{sec:confirmation}.
The initial observation of G141.2+5.0 used 7 TABs in total, that overlap at $0.5\times$\,FWHM. We folded the other six beams on the given period and DM, but no signal was seen. \refbf{Given the low S/N of the candidate in the central beam, that non-detection is consistent with a radio signal from the central X-ray point source, as this signal would be further attenuated in these outer beams,  }
To verify that the signal was not an instrumental artifact we performed a search for periodic signals at a DM of 226.00 \dm. No candidates of interest were found. 

In other observations, no candidates were detected.
For these we derive upper limits below. 

\subsection{Derived upper limit of PSR J1931+30}
The line-of-sight to SNR G65.3+5.7 harbours the pulsar J1931+30 \citep{camilo1996}. In our observation it is  located at the intersection (and FWHM) of beam 15 and 16, see Fig.~\ref{fig:beams}. We did not find this pulsar in our blind search. 
Searching beam 15 and beam 16 more thoroughly, assuming a spin-down rate between $10^{-11}$ and $10^{-14}\, \rm{s\,s^{-1}}$ at its DM of $56\pm5$ \dm also did not yield in any detection of this source, resulting in an upper limit at 150\,MHz of $6\pm3$\,mJy \refbf{(see Sect. \ref{sec:sensitivity})}.
That agrees with a spectral index of $\alpha=-1.4$ or shallower.

\subsection{Sensitivity limits}
\label{sec:sensitivity}
\subsubsection{Sky background}
The sky temperature $T_\mathrm{sky}$ is the dominant contributor to the  \mbox{LOFAR} HBA  system temperature $T_\mathrm{sys}$.
Generally this sky background consists mostly of the diffuse synchrotron emission from our Galaxy.
In any survey for pulsars in supernova remnants, care needs to taken to include the nebular emission.
This is a significant effect in e.g. Crab pulsar observations with LOFAR \citep{bilous2016, lkm+18}.
Our general sky noise map is scaled from the \citet{hssw82} 408\,MHz map.
Remnants G65.3+5.7 and G150.3+4.5 are larger than the beam sizes used in this survey, and are resolved. 
Similarly, G49.2$-$0.7 and G93.3+6.9 are of comparable size to the map resolution. 
The noise effect of these \refbf{four SNRs is thus included in the Haslam map, and through there, in  our pulsar search sensititivy estimates.}
For the other, smaller remnants we add an extra sky temperature component. 
\refbf{To be conservative we base this on the addition that G63.7+1.1, that largest of the remaining four, makes to the background.} 
\citet{wallace1997} show the spectrum of G63.7+1.1 is quite flat, and from their 327\,MHz measurement of 1.5\,Jy,
scaling as $\nu^{-0.3}$ (very similar to the $-$0.27 power-law seen in the Crab Nebula; \citealt{1997ApJ...490..291B}) 
we estimate the 150\,MHz SNR flux to be $1.9 \pm 0.2$\,Jy.
\refbf{In our observations, however, that integrated flux density is divided over 7 beams.} 
\refbf{This adds 0.3\,Jy to the time-domain background, equivalent to about 1\,K.} 
\refbf{That is not a significant addition to our system noise of 10$^4$\,K, but for completeness} 
we have included it in our sensitivity estimates below.
\subsubsection{Sensitivity estimates}
\label{sec:sens}
We base our upper limits on the peak S/N of candidate \cand. 
\refbf{If the pulsar beam is so narrow that the profile is a single bin, then peak and integrated S/N are identical. We quote the peak S/N as the profile broadness is unknown in a blind search.}
The peak signal-to-noise ratio for the profile shown top-left in Fig.~\ref{fig:cand} is 4 (the integrated profile S/N is 10) and corresponds to an integrated-profile flux-density of $7.6\pm3.8$\,mJy.
Based on this peak S/N, we report 4 $\times$ noise-rms flux-density, $\sigma_{\rm{RMS}}$, as our upper limit for the mean flux-density of the other sources.
By necessity this is a very one-dimensional metric. Traits such as the signal consistency throughout the observation, and signal visibility over the entire bandwidth are not contained in it, but important during human inspection.
The sensitivity limits can be found in Table~\ref{table:data_reduction}.

We obtain $\sigma_{\rm{RMS}}$ following the procedure\footnote{Implemented in python as \url{https://github.com/vkond/LOFAR-BF-pulsar-scripts/blob/master/fluxcal/lofar_fluxcal.py}} described in \cite{kondratiev2016}. It derived the telescope noise levels from the radiometer equation for the exact observations parameters such as zenith angle, and the effective area of each individual included station.

In Table \ref{table:data_reduction} we report the sensitivity limits for the coordinates of the central beam of each observation.
The LOFAR sensitivity in a given observation  depends strongly on elevation, and its influence on the effective area.
This is visible from the source declination and beam-size in Table \ref{table:sources} and our reported $S_{\rm{min}}$ in Table \ref{table:data_reduction}.
That dependence plus the fact that only 20 out of 23 core stations were available during the observation of G49.2$-$0.7, results in a poorer sensitivity for that observation.
For completeness we check the gradient of this sensitivity fall-off with elevation over our more extended objects.  Even in the largest source the sensitivity variation is within $3\%$ of central beam.
Finally, the reduced sensitivity towards G65.3+5.7 and G150.3+4.5 is a result of the trade-off between sensitivity and field of view, to maximise survey speed. For these large sources using only the Superterp (inner 6) stations with their large tied-array beams, provided best sensitivity over the entire remnant within the allocated observing time.

\begin{figure}
\centering
\includegraphics[width=\columnwidth]{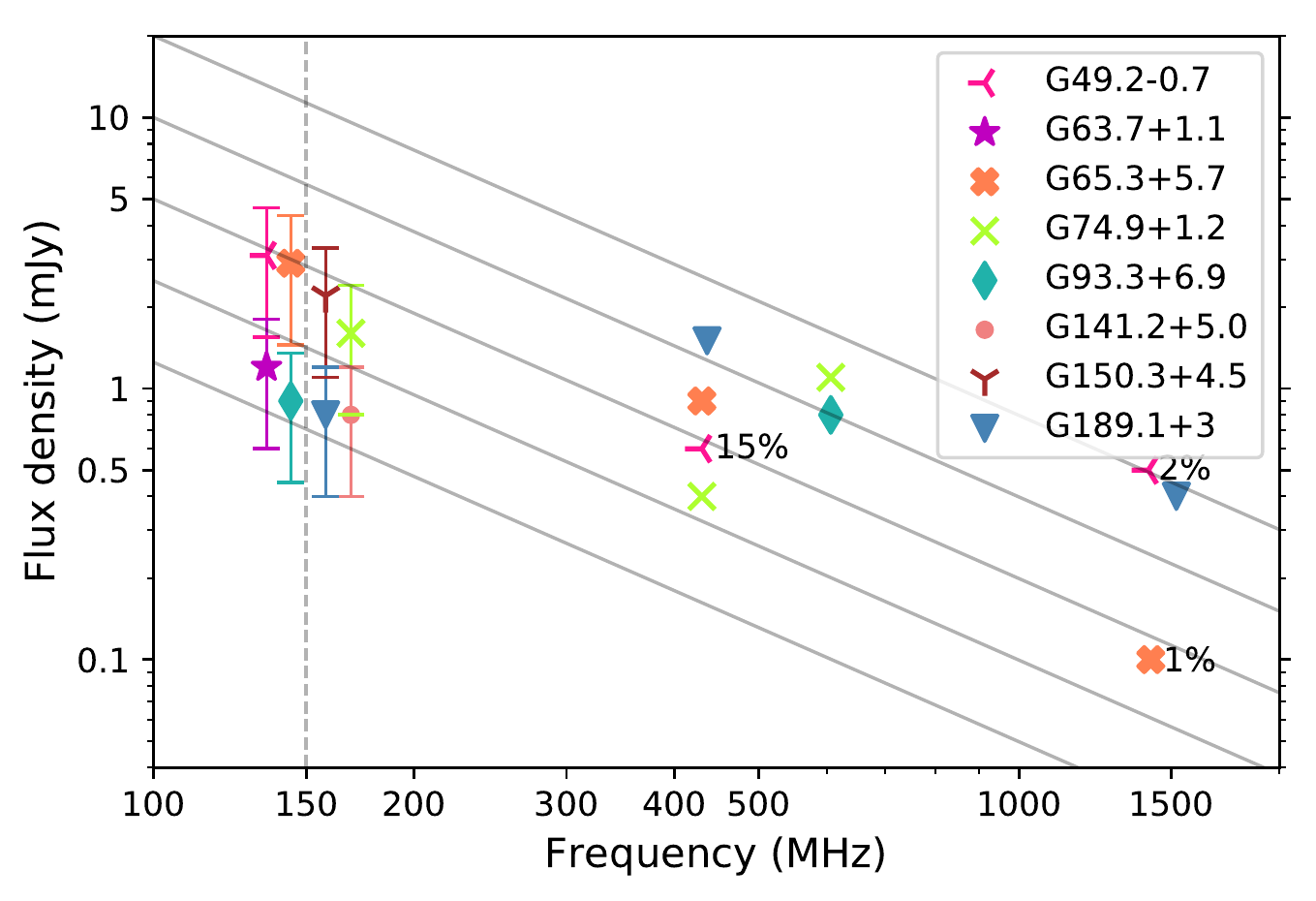}
\caption{
Our  sensitivity limits (left, at 150\,MHz) compared to applicable earlier work.
For previous searches that did not cover the entire supernova remnant size, the searched fraction is indicated. Per frequency we only show the most sensitive limits obtained.
We overlay a grid with spectral slope $\alpha = -1.4$ to indicate the scaling of our sensitivity limits \citep{bates2013}. 
For visibility our limits are plotted slightly offset from the 150\,MHz observing frequency (dashed line).
\label{fig:sensitivity_limits}
}
\end{figure}

\begin{table*}
\caption{The expected DM
upper limit on the mean flux density $S_{\rm{min}}$, and maximum possible pseudo luminosity L$_{\rm{max}}$ for each of the sources. DM models are NE2001 \citep{cordes2002} and YMW16 \citep{ymw16}.
The errors on the DM are based on distance error (see Table~\ref{table:sources}). In the case of SNR G150.3+4.5 where no distance is known, the maximum galactic DM for its line-of-sight is given.  
The error on LOFAR flux density estimates, and thus on our $S_{\rm{min}}$, is 50\%.
\label{table:data_reduction}
}
\centering
\begin{tabular}{l r r r c c}
\hline\hline
Source & DM NE2001      & DM YMW16       & DMs searched & $S_{\rm{min}}$ & $L_{\rm{max}}$ \\
       & (\dm)          & (\dm)          & (\dm)        & (mJy)          & (mJy kpc$^2$) \\
\hline
G49.2$-$0.7 & $99.43\substack{+38.5 \\ -36.9}$ & $132.72\substack{+30.02 \\-26.39}$ & $10-400$ & $3.1\pm1.5$ & $ 90 \pm 44 $	 \\
G63.7+1.1 & $73.39\substack{+57.35 \\-46.53}$ & $61.61\substack{+21.09 \\-25.35}$ & $30-400$   & $1.2\pm0.6$ & $ 40 \pm 20 $	 \\
G65.3+5.7 & $10.04\pm 2.16$ & $13.33\substack{+2.65 \\-2.67}$ & $2-200$                        & $2.9\pm1.4$ & $ 2.3\pm 1.1 $    \\
G74.9+1.2 & $178.46\substack{+45.43 \\-44.28}$ &  $197.67\substack{+71.44 \\-87.92}$ & $2-200$ & $1.6\pm0.8$ & $ 60\pm 29 $    \\
G93.3+6.9 & $34.11\substack{+15.54 \\-14.27}$ & $37.42\substack{+11.37 \\-9.09}$ & $10-350$    & $0.9\pm0.5$ & $ 11 \pm 6 $	 \\
G141.2+5.0 & $123.21\substack{+14.31 \\-17.17}$ & $147.00\substack{+5.55 \\-5.70}$ & $30-400$  & $0.8\pm0.4$ & $ 13 \pm 6 $	 \\
G150.3+4.5 & $\leq180$ & $\leq260$ & $10-400$                                                  & $2.2\pm1.1$ & $  - $	         \\ 
G189.1+3 & $82.04\substack{+21.03 \\-18.13}$ & $121.78\substack{+26.08 \\-40.08}$ & $20-400$   & $0.8\pm0.4$ & $ 1.8 \pm 0.9 $   \\
\hline
\end{tabular}
\end{table*}

\section{Confirmation Observing and Data Reduction}
\label{sec:confirmation}
Under LOFAR Directors Discretionary Time (code: DDT6\_001), our first attempt to confirm candidate \cand was scheduled on
2016 June 13, when we re-observed the central beam of G141.2+5.0 (see Fig.~\ref{fig:beams}).
Data from 23 core stations were coherently beam formed.
The increased observation duration of 5h15m  
improved our sensitivity by 33\% over the limits sets in search mode (0.6\,mJy vs. 0.8\,mJy, see Section \ref{sec:discussion}).
We performed a search over the parameter space spanned by the unknown spin-down.
In this search we also accounted for a change in DM of $\pm\,20$\,\dm due to potential filamentary structure in the PWN. 
Yet the pulsar was not detected.

On 2017 January 8, in our second confirmation attempt, 
we combined the same stations. 
For 6 hours we now recorded the beam-formed 
data in complex-voltage mode \citep[cf.][]{stappers2011}.
We wrote these baseband data to disk such that we could coherently dedisperse the data off-line. 
In the initial detection, the apparent pulse width $w \simeq 6$\,ms (Fig.~\ref{fig:cand}).
Given the relatively high dispersion measure of the candidate, 226\,\dm, 
up to 3.3\,ms of this could be instrumental broadening
caused by intra-channel dispersion smearing.
Coherent dedispersion eliminates this factor.
If we follow the LOFAR radiometer equation  \citep[Eq.~3 in][]{2010A&A...509A...7V} for the pulse period $P=94$\,ms, initial pulse width $w \simeq 6$\,ms and assumed follow-up width \mbox{$w \simeq 4$\,ms},
the resulting more concentrated pulse translates to a 20\% increased sensitivity (cf.~Sect.\ref{sec:sens}).
From the increased observation duration we also derive an $S_{\mathrm{min}}=0.55$\,mJy, which is 40\% more sensitive than the initial detection.
When both combined, our overall sensitivity increased by as much as $50\%$.

On recording, the 78\,MHz of 8-bit baseband data were split over 20 subbands.
These 6\,TB were stored from the Poznań Supercomputing and Networking Center branch of the LTA (from where they are publicly available), and transferred to Cartesius.
There, data were coherently dedispersed within each subband using {\tt CDMT\footnote{\url{https://github.com/cbassa/cdmt}}} \citep{2017A&C....18...40B}.
Subbands were combined to coherently dedispersed filterbank files containing 3200 channels at 164\,$\mu$s sampling. 
After removing radio frequency interference,
candidates were next folded over a range around  their best frequencies,
and blindly searched over a wider set of frequencies.
No pulsar signals were seen.  
With the same setup, test pulsar J0218+4232 was easily detected at its DM and period.

In summary, we were unable to confirm candidate \cand.

\section{Discussion}
\label{sec:discussion}

\subsection{Candidate pulsar \cand in G141.2+5.0}
Due to the modest significance of candidate \cand we cannot rule out that non-guassian noise processes (from RFI or instrumental) played a role in fortuitously mimicking a pulsar-like signal. 
And the large number of trials in our search gives ample opportunity for such a chance event.

Still, a dozen other LOFAR pulsars this weak or weaker have been successfully confirmed.
What could have caused the initial detection but the later confirmation failures?

\subsubsection{Scintillation}
Pulsar signals are prone to several propagation effects caused by the ionized interstellar medium (IISM). \refbf{Temporal variations of the dispersive delay and interstellar scattering observed in other pulsars are insufficient to weaken the candidate signal on a one year timescale \citep{petroff2013,hemberger2008}.}
However, G141.2+5.0 is identified as a PWN \citep{reynolds2016} and PWNe are known to be highly turbulent media and show filamentary structure.
The line-of-sight medium could have changed in between our observing epochs, enhancing the scattering properties and causing the already weak signal to be below our detection threshold.

A propagation effect which \emph{is} known to vary on several time-scales is scintillation. In the observed frequency  range the IISM causes propagation in the strong regime where the scintillation can be subdivided in refractive and diffractive scintillation. 
Diffractive interstellar scintillation (DISS) causes intensity variations in both time and frequency. The decorrelation bandwidth can be approached by Eq.~\ref{eq:decorrelation} \citep{cordes1985}:

\begin{equation}
\label{eq:decorrelation}
\Delta f_{\mathrm{DISS}} \sim 11\, \mathrm{MHz} \left(\frac{f}{\mathrm{GHz}}\right)^{4.4}\,\left(\frac{d}{\mathrm{kpc}}\right)^{-2.2}
\end{equation}
For G141.2+5.0 the decorrelation bandwidth of 0.123\,kHz is smaller than the used channel width (6.1 kHz), ruling out diffractive scintillation. 
However, refractive scintillation (RISS) operates on longer time-scales (e.g., $\sim$ months to years) and allows for a change in observed pulse intensity as demonstrated by \cite{stinebring2000}. 
One of the possible reasons  we do not re-detect the initial pulse signal can be that RISS decreases the flux density to below our observing thresholds.
\refbf{In that case, the 14 months between the initial detection and first confirmation attempt must be the appropriate RISS timescale. Confirmation observations spaced with a similar yearly cadence can test this hypothesis.}

\subsubsection{Plasma Lensing}
Noteworthy also, is the overlap with the low rate of re-detections in Fast Radio Bursts.
Those sources too -- at least some -- are thought to reside in supernova remnants \citep{Straal-2018}.
Plasma lensing in that host environment may sometimes boost up the signal strength. At those times the FRB is detected.
Yet follow-up observations of even the most high-significance FRBs has been fruitless \citep{2015MNRAS.454..457P}.
Potentially the same effect is seen, much more nearby, in candidate \cand.

\subsubsection{Intermittency}
\refbf{Among the handful currently known intermittent pulsars \citep{2017ApJ...834...72L} some show  ON fractions as low as 0.8\%.}
Since most research on intermittent pulsars is performed at higher frequencies (e.g. 1\,GHz), 
discovering a low-frequency intermittent pulsar would certainly be of interest.
\refbf{Furthermore, while the characteristic age of the five known intermittent pulsars is below the median age of the entire population, none of them are younger than 10$^5$\,yr. Finding a young pulsar with intermittent behavior could potentially shed further light on the magnetospheric circumstances conducive to such mode switching, as the short period creates different voltage gaps and currents.}
Given our continued confidence in the initial detection, intermittent behavior is an option.
But confirming  such an intrinsically variable source is  challenging.

\refbf{
\subsubsection{Future confirmation options}
In three LOFAR observations of increasing sensitivity,  candidate \cand was only seen in the first session. It was not blindly detected at 820 MHz with GBT; and as data from that session was not archived, it could not be refolded specifically at the candidate period. Now that the DM is known, coherent dedispersion can be applied in confirmation observations; and the observed scatter broadening is low (cf.~\S\ref{sec:datareduction}). These two factors mitigate the major disadvantages of low-frequency searching. 
At 150\,MHz, a 2019 repeat observation using  LOFAR could be confirm the source if scintillation has since amplified it.
At 350\,MHz, a deep observation with the GBT could potentially be constraining. For a spectral index $\alpha$=$-1.4$, a 2-hr integration can reach a limit about 2$\times$ deeper than reported here (comparing e.g. \citealt{2011A&A...531A.125C} to Table \ref{table:data_reduction}).
At much higher energy, the detection of the 94-ms periodic signal in the X-rays emitted by the known image point source would confirm the candidate.  
Assuming a 50\% pulse fraction, \textit{XMM Newton} timing mode observations are capable of this measurement in 25\,ks, but no observations have so-far been carried out.
}

\subsection{Non-detections -- Upper limits}
In Fig.\ref{fig:sensitivity_limits} we compare the  sensitivities obtained in our searches
to  previous surveys, conducted at higher frequencies, often covering only a fraction of the central remnant area.
To properly compare our sensitivities we take into account the average pulsar flux-density spectral slope of $\alpha=-1.4$ (see Sect. \ref{sec:introduction}). 
After applying this scaling, our limits on one source, G74.9+1.2 are roughly of equal sensitivity as  previous searches. 
\refbf{Scaling the 430\,MHz observation results in an expected flux density at 150\,MHz of 1.9\,mJy vs. 1.6\,mJy obtained in this work.}
For all other sources searched before (G49.2-0.7, G65.3+5.7, G93.3+6.9, and G189.1+3), we obtain full coverage while \emph{also} obtaining improved limits by as much as a factor of 10 (G189.1+3). 
Also for sources on which our searches are the first, we obtain very stringent sensitivity limits.

However, except for G141.2+5.0, as discussed above, we do not find any candidate in the other observations.
These non-detections can have multiple reasons, which we discuss below.
We order them starting from the most basic and pulsar centric -- its formation -- 
all the way to the hurdles for the final detection of its signal, by  beaming and propagation effects.

\subsubsection{Neutron star vs. black hole formation}
It is possible that in stead of a pulsar, a non-emitting neutron star or even a black hole is formed.
\refbf{Some radio-quiet neutron stars are observable as CCOs in the high-energy band, as mentioned in the introduction.}
The detection of an X-ray point source in G74.9+1.2 and G189.1+3 are indications that, at least there, a neutron star is formed.
\refbf{Yet, both these sources have a PWN, indicating the presence of an active rotation-powered pulsar instead of a CCO.
The two SNRs that do not host a PWN (see Table \ref{table:sources}) may have formed a CCO. 
Our stringent upper limits then argue in favor of CCOs as a radio-quiet class.
}

However, for the five sources without a confirmed point source, it is possible that the star collapsed to a black hole.
As mentioned in Sect. \ref{sec:introduction}, it is expected that between 13\% and 25\%  of the cc-SNe produce black holes \citep{heger2003}. Of the 383 SNRs reported in the online SNR catalogue (June 2018), we can expect that 75\% are the result of a cc-SN \citep{cappellaro1999}. Of those 287, 170 SNRs have a (possible) NS, pulsar or PWN association. This is 60\% of the cc-SNRs.
For simplification we assume that a PWN implies the presence of a pulsar. 
So in two thirds of the remaining SNRs it is possible that a black hole is formed. Previous research argues that SNR G93.3+6.9 is the result of a sub-energetic SN and has likely formed a NS \refbf{\citep{jiang2007}}.
The other three sources in our sample, without associated PWN, could very well harbour a black hole rather than a pulsar.

\subsubsection{Transversal velocity}
In the SN explosion, pulsars receive a natal kick allowing them to travel away from their birth site. 
For those sources without an X-ray point source this could mean that the pulsar has travelled away from the centre, or even out of the remnant.
As we searched the full area of each remnant without a point source, travelling out of the centre is insufficient to explain the non-detection. Here, the pulsar must have completely left the remnant.
We find that, using the parameters on SNR size and age provided in Sect. \ref{sec:sources}, 
this requires transversal velocities exceeding 750\,km\,s$^{-1}$.
The mean transversal velocity for  single radio pulsars preferred by \cite{faucher2006}, 
 of $180\substack{+20\\-30}$ km\,s$^{-1}$,  translates to a 3D velocity of: $v_{\rm{3D}} = 380\substack{+40\\-60}$ km\,s$^{-1}$. This is well below the velocity required for escaping the remnants. 

Such high space velocities would produce bow shocks, which are not observed.
We therefore conclude that the pulsars are still contained in their parent environment, and covered by our beam layout.

\subsubsection{Beaming fraction}
A necessary and important condition to be able to detect even the brightest pulsar is that its radio beam sweeps over us.
The part of the sky, as seen from the frame of the pulsar, that the radio beam covers, defines the \emph{beaming fraction}. 
This fraction thus also describes the chance the source is visible from Earth.
The directly measurable decreasing pulse width for longer-period, older pulsars already suggests a trend from large beaming fractions for young pulsars to small fractions late in life \citep{1988MNRAS.234..477L}.
The recently discovered, old, 24-s pulsar {J0250+5854} displays a duty cycle of only 0.3\% \citep{tan2018}.
The evidence is not just anecdotal -- when looking at the entire pulsar population, the dearth of older pulsars also means their beaming fractions decrease until few are visible \citep{lv04}. Young pulsars in this population, however, must have high beaming fractions. 

Still, young pulsars do not regularly show the  broad profiles one would expect for circular beams with high beaming fractions. A different noteworthy characteristic of their profiles is, however, an overabundance of interpulses. These already suggest the radio beams in young pulsars are elongated in latitude. The first hints seen in a relatively small sample of polarization measurements by \citet{1983A&A...122...45N} were recently confirmed in \refbf{PSR~J1906+0746,  where 
geodetic precession, caused by the binary interaction with its companion \citep{2015ApJ...798..118V},
allowed for mapping of the main and interpulse components; and  indeed significantly elongation is observed \citep{dkc+12}.}

Studies of beaming fractions were often complicated by our inability to determine the "ground truth"; 
what is the actual count of emitting sources, over the number we see.
Beaming fractions could be derived by \emph{starting} from known SNRs and determining how many pulsars 
were detected \citep[cf.][]{2000ASPC..202..485K}. But including any such results here would circularize our reasoning. 
The central source of a PWN is beyond question, so these can be confidently used as ground truth.
From  a study of such nebulae, \citet{1993ApJ...408..637F} find a beaming fraction of $0.6\pm0.1$.
More recently,  the increasing number of $\gamma$-ray detected pulsars offers a new angle of analysis. 
In a study comparing the $\gamma$-ray and radio detection probabilities of young pulsars,
\citet{2010ApJ...716L..85R} find the radio beaming fraction of young, energetic pulsars 
is $>$0.6, and most likely approaches unity. 

Thus, given an unbiased sample of pulsars in SNRs, the beaming fraction of young pulsars is high enough to confidently rule this out as cause for non-detections in all surveyed sources. Yet part of the 8 in this study were chosen because no pulsars had been detected yet; if we look at the total population of $\sim$60 SNR pulsars, with a beaming fraction of 0.9 it is possible that the 8 in our study are the subset whose beams miss us.

\subsubsection{Luminosity}
\label{sec:lum}
Given the known luminosities of young pulsars, could these sources be beamed at us but too dim to detect?
We first compare our set to the Crab pulsar. The target farthest from Earth is 
G74.9+1.2 at $6.1 \pm 0.9$\,kpc. 
Moving, for sake of argument, the Crab pulsar there from its current distance of 2\,kpc would decrease the flux density by an order of magnitude. Yet, within an hour, LOFAR observes multiple Crab giant pulses of S/N of several hundred \citep{lkm+18}. These would thus easily be detectable for all SNRs in our set. The periodic signal of the Crab, with its mean flux density of $8 \pm 4$\,Jy at LOFAR frequencies \citep{bilous2016}, would also be far above our sensitivity limits as listed in Table \ref{table:data_reduction}, even after being diluted by a maximum factor of 10 by the inverse square law.

We next aim to compare our limits to the luminosity of the ensemble of known young pulsars. 
For this we use the pseudo-luminosities $L_{\rm{max}}$ =  $S_{\rm{min}}$ $\times$ d$^2$ presented in Table~\ref{table:data_reduction}.
However, of the 81 pulsars in the ATNF catalog\footnote{\url{http://www.atnf.csiro.au/research/pulsar/psrcat}} \citep{manchester2005} having a pulse period $P$ less than 100\,ms and a characteristic age $\tau$ below 100\,kyr, just 13 have a 400-MHz flux density measurement. Only from these values can a somewhat dependable 150-MHz flux density be estimated. The other 68 have reported fluxes only at 1.4\,GHz and higher. Extrapolating those down over an order of magnitude in frequency would introduce to much uncertainty, given the variations in spectral index.

For these 13, pseudo luminosities at 150\,MHz are expected to be in the range of $\sim10-10^4$ mJy kpc$^2$. For our observation with the least stringent $L_{\rm{max}}$ limit of 90 mJy\,kpc$^2$, toward G49.2$-$0.7, half of these 13 would have been detected. In our observations of G74.9+1.2 and G63.7+1.1, two thirds of the sample would be found. In all other observations, each of the existing pulsars would have been detected, save for an occasional miss on very dim source PSR~J0659+1414 \citep{2010ApJ...708.1426W}, which is only 9\,mJy\,kpc$^2$.

\refbf{The uncertainty introduced by scaling 400-MHz fluxes down to 150\,MHz could be avoided by using only the LOFAR fluxes reported in \citet{bilous2016}. But among these only young two pulsars ($P < 100\,ms$ and $\tau$ < 100\,kyr; or associated with a SNR) are found, PSR~B0531+21 and PSR~B0656+14. Still, a statistical comparison by means of a Kolmogorov–Smirnov test indicates that the 150-MHz pseudo luminosities from the \emph{entire} \citet{bilous2016} sample of 152 pulsars are indistinguishable from those extrapolated for the set of 13 young ATNF pulsars. The spectral index uncertainty thus does not play a significant role on our completeness estimate.}

Given the known pulsar luminosities, we conclude our searches were sensitive and deep enough for all targets save perhaps, by chance, one.

\subsubsection{Propagation effects}
Finally, it could be that a pulsar is formed, which resides in the remnant and is beamed towards us, and bright enough in principle, but still undetectable due to e.g., pulse broadening.
As the impulsive signal travels through the IISM it gets smeared out in time by dispersion and scattering.
The dispersion of the signal scales with $\nu^{-2}$ and therefore is stronger at lower frequencies.
This limits the DM range we are able to search. Signals with a very high DM would be too broadened  for us to detect.
At our observing frequency, overly high DMs cause intra-channel dispersion smearing which reduces our S/N 
(cf. the discussion on the confirmation observations presented in Sect.~\ref{sec:confirmation}).
In our observational set-up this effect becomes noticeable above DMs of 100 \dm.

Our source selection criteria aimed to limit the expected DMs, based on the distance information from the literature. 
Potentially, some of these distances and DMs were underestimated.
For individual sources the NE2001 model can be off by a factor $1.5-2$ \citep{schnitzeler2012}. 
Additionally, in \citealp{Straal-2018} we show that pulsars in SNRs and PWNe generally exhibit 
DMs that are 20 \dm higher than the model predicts.

A further propagation effect that highly affects the low-frequency signal is scattering.
Scattering scales with $\nu^{-4.4}$ for a Kolmogorov turbulent medium and is therefore more prevalent at lower frequencies. The multi-path propagation of the signal causes pulse smearing for which we cannot correct in the data reduction.

Although we have taken pre-cautions in our source selection, both propagation effects can highly affect the signal strength; and could explain the non-detections.

\refbf{Higher-frequency observations suffer from these deleterious effects less, but on these extended sources that advantage is generally offset by the smaller field of view of the  telescopes that offer such frequencies.
}

\section{Conclusions}
\label{sec:conclusions}
We have performed the first low-frequency directed pulsar search towards eight selected SNRs and PWNe.
Some of these had previously  been searched at higher frequencies.
We find one very interesting pulsar candidate, \cand, towards G141.2+5.0 at the position of the X-ray point source, with a period of 94.13\,ms and at a DM of 226 \dm.
We re-observed the source twice, with a 30\% and 50\% increased sensitivity, but were unable to confirm the candidate. 
Its low detection S/N of 4 could mean our fortuitous initial detection and subsequent non-detections in the follow-up were due to enhanced scattering or refractive scintillation.

One non-associated pulsar in the line of sight to G65.3+5.7, PSR J1931+30, was located at the edge of two of our observing beams, and not detected.
Our upper limit of $6\pm3$\,mJy at 150\,MHz, agrees with a spectral index of $\alpha=-1.4$ or shallower.

For all sources we obtain improved sensitivity limits and we discuss the reasons of the non-detections. 

Two unique aspects of this SNR/PWN pulsar search, the largest such low-frequency survey to date, are the wide field, and the high sensitivity. 
These allow us to conclude the following: First, from our wide field we determine that the putative pulsars can not have left their remnants. 
Second, from our high sensitivity we infer that some of these SNRs do not host a neutron star, but a black hole; 
and that the remainder is invisible due to propagation and beaming effects.

\begin{acknowledgements}
The research leading to these results has received funding from the European Research Council under the European Union's Seventh Framework Programme (FP/2007-2013) / ERC Grant Agreement n. 617199, and from the Netherlands Research School for Astronomy (NOVA4-ARTS). This work was carried out on the Dutch national e-infrastructure with the support of SURF Cooperative. Computing time was provided by NWO Physical Sciences (project 15236). \\
We thank \citet{KapteynPackage} for developing the Kapteyn Package and \citet{psrqpy} for {\tt psrqpy}. \\
This paper is based on data obtained with the International LOFAR Telescope (ILT) under project codes LC3\_024, DDT6\_001 and LC7\_032. LOFAR \citep{2013A&A...556A...2V} is the Low Frequency Array designed and constructed by ASTRON. It has observing, data processing, and data storage facilities in several countries, that are owned by various parties (each with their own funding sources), and that are collectively operated by the ILT foundation under a joint scientific policy. The ILT resources have benefited from the following recent major funding sources: CNRS-INSU, Observatoire de Paris and Université d'Orléans, France; BMBF, MIWF-NRW, MPG, Germany; Science Foundation Ireland (SFI), Department of Business, Enterprise and Innovation (DBEI), Ireland; NWO, The Netherlands; The Science and Technology Facilities Council, UK; Ministry of Science and Higher Education, Poland. 

\end{acknowledgements}

% for the bibliography, at the end 
\bibliographystyle{yahapj} % style aa.bst 
\bibliography{Straal_vanLeeuwen_ms}
\end{document}